# A Unified Frequency-Domain Model for Cascaded Filter-Interpolation Modulation in Tomographic Reconstruction

Detian Li[1], Yang Zou[1], Penghao Geng[1], and Shengkun Yao[1,*]

[1]Shandong Provincial Key Laboratory of Light Field Manipulation Physics and Applications &School of Physics and Optoelectronics, Shandong Normal University, Jinan 250358, China

*yaoshk@sdnu.edu.cn

**The fidelity of image reconstruction from projections in linear inverse problems, such as tomography, is critically dependent on the synergistic interaction between frequency-domain filtering and spatial-domain interpolation. However, a physical model that can quantitatively describe how these two components cascade interact in the frequency domain and ultimately determine image quality is still lacking to this day. Here, we introduce a unified frequency-domain model that conceptualizes the combined effect of filtering and interpolation in the filtered backprojection (FBP) algorithm as a cascaded modulation process. This model demonstrates that the effective reconstruction spectrum is determined by the original projection data being sequentially modulated by the frequency responses of the filter and the interpolation kernel. Comprehensive numerical simulations and synchrotron radiation CT experiments validate the model, confirming its power to explain the performance hierarchy of classical filter-interpolation pairs under both ideal and noisy conditions. The model successfully predicts key performance characteristics, including spatial resolution and structural fidelity, thereby elucidating the physical principles behind**

**the efficacy of specific combinations. This work establishes a generalizable theoretical foundation for analyzing cascaded systems in linear inverse problems, moving the practice of algorithm selection in computational imaging from empiricism to a principled, physics-based paradigm.**

The inverse problem of reconstructing an object's internal structure from its line-integral projections is foundational to numerous imaging modalities[1,2]. As a solution to this problem, tomographic reconstruction provides non-invasive, multi-dimensional visualization capabilities that have revolutionized fields from medical diagnostics to materials science[3-6]. The fidelity of reconstructed images is intrinsically governed by the reconstruction algorithm.

Among diverse reconstruction techniques, the Filtered Backprojection (FBP) algorithm remains a cornerstone method widely deployed in clinical and industrial settings[7-9]. Its prevalence is attributable to computational efficiency, straightforward implementation, and its rigorous foundation in the Fourier Slice Theorem[10]. The FBP process inherently decomposes into two distinct operations: frequency-domain filtering that compensates for inherent blurring from simple backprojection[11], and spatial-domain backprojection with interpolation that maps discretely sampled projections onto a continuous image grid[12].

Extensive research has been devoted to independent optimization of FBP's two core components. Various filters have been developed, including Ram-Lak, Shepp-Logan, and Hann filters[13-17]. Concurrently, multiple interpolation kernels have been explored, such as Nearest-neighbor, and Linear methods[18-21]. However, in practical FBP implementation, the filter and interpolation operator form a cascaded system where their frequency responses engage in multiplicative interaction that collectively sculpts the final system transfer function and dictates reconstructed image quality[22].

Despite the fundamental importance of this synergistic interaction, a rigorous physical model quantitatively describing this cascaded modulation process remains underdeveloped[23]. The empirical approach of testing combinations posteriori fails to provide first-principles understanding of why certain filter-interpolation pairs synergize to produce superior results while others lead to antagonistic outcomes. This knowledge gap hinders efficient parameter selection in tomographic workflows and limits fundamental understanding of signal chains in linear inverse problems.

To bridge this gap, we introduce a unified frequency-domain model that formally characterizes the interpolation-filtering interaction in FBP as a cascaded modulation process. This model posits that the effective reconstruction spectrum is governed by the product of the filter frequency response and the interpolation kernel spectrum, providing a quantitative and predictive theory for final image quality. Through systematic investigation combining numerical simulations with synchrotron CT experiments, we comprehensively evaluate classical filter-interpolation combinations, empirically validating the model. Our work moves beyond isolated parameter tuning, establishing a first-principles framework for understanding, predicting, and optimizing this critical physical process in tomographic image formation.

The FBP algorithm provides an efficient solution to the inverse problem of recovering a two-dimensional attenuation coefficient distribution $f(x, y)$ from its one-dimensional line integrals[24]. This process is governed by the Fourier slice theorem [25]. Let $I$ be an arbitrary straight line along a ray $L$ in the plane, q denotes the filtered projection, and $\theta$ represent the projection angle. The integral of $f(x, y)$ along the line $I$ is referred to as the Radon transform. The formula for FBP can be expressed as:

$$f(x, y) = \int_0^{\pi} q(x\cos\theta + y\sin\theta, \theta)d\theta, \quad (1)$$

Physically, the ramp filter $|\omega|$ compensates for the inherent $1/|\omega|$ blurring introduced by the simple backprojection process. The filtering phase is therefore essential for reversing this blur and recovering high-frequency content. At the same time, the choice of interpolation method, integral to the discrete implementation of the backprojection operation, significantly affects the geometric fidelity and spatial accuracy of the reconstructed image. Their combined action ultimately governs key image quality.

Filters operate by applying a frequency response function $H(\omega)$ to weight the projection data spectrum $P(\omega,\theta)$, thereby controlling the gain of different spatial frequency components ω to achieve a trade-off between image resolution and artifact suppression.

Let $\omega$ denote the spatial frequency and $\omega_0$ the cutoff frequency. The frequency response function of the Ram-Lak filter is defined as[26]:

$$H_{\text{Ram-Lak}}(\omega) = |\omega| \text{rect}\left(\frac{\omega}{2\omega_0}\right) = \begin{cases} |\omega|, & \text{if} |\omega| \leq \omega_0 \\ 0, & \text{otherwise} \end{cases}, \tag{2}$$

Let $B$ denote the bandwidth. The frequency response of the Shepp-Logan filter is defined as[27]:

$$H_{\text{Shepp-Logan}}(\omega) = |\omega| \text{sinc}\left(\frac{\omega}{2B}\right), \tag{3}$$

Where $|\omega| \leq B$.

The frequency response of the Cosine filter is defined as[28]:

$$H_{\text{Cosine}}(\omega) = |\omega| \, cos\left(\frac{\pi\omega}{2B}\right), \tag{4}$$

The frequency response of the Hamming filter is defined as[29]:

$$H_{\text{Hamming}}(\omega) = |\omega| \left[0.54 + 0.46 \, cos\left(\frac{\pi\omega}{B}\right)\right], \tag{5}$$

The Hann filter utilizes a Hann window function that decays rapidly to zero. The frequency response of the Hann filter is defined as [30-31]:

$$H_{\text{Hann}}(\omega) = |\omega|\frac{1+cos\left(\frac{\pi\omega}{B}\right)}{2}, \tag{6}$$

The magnitude responses $|H(\omega)|$ of different filters within the normalized frequency range are depicted (Fig. 1). The Ram-Lak filter shows a sharp rise in low frequencies, dropping abruptly to zero at cutoff. It preserves high-frequency details well but is highly noise-sensitive. The Shepp-Logan offers a smoother response, balancing detail enhancement and noise suppression. The Cosine filter decays smoothly in a symmetric cosine shape, providing moderate noise reduction. Both Hamming and Hann are symmetric low-pass filters: Hamming has a wider main lobe and lower sidelobes, better preserving mid-frequencies, while Hann suppresses high-frequency noise more effectively due to faster sidelobe decay. Without filtering, the response remains constant, causing severe artifacts and noise. This figure aids in selecting filters based on data features and reconstruction needs.

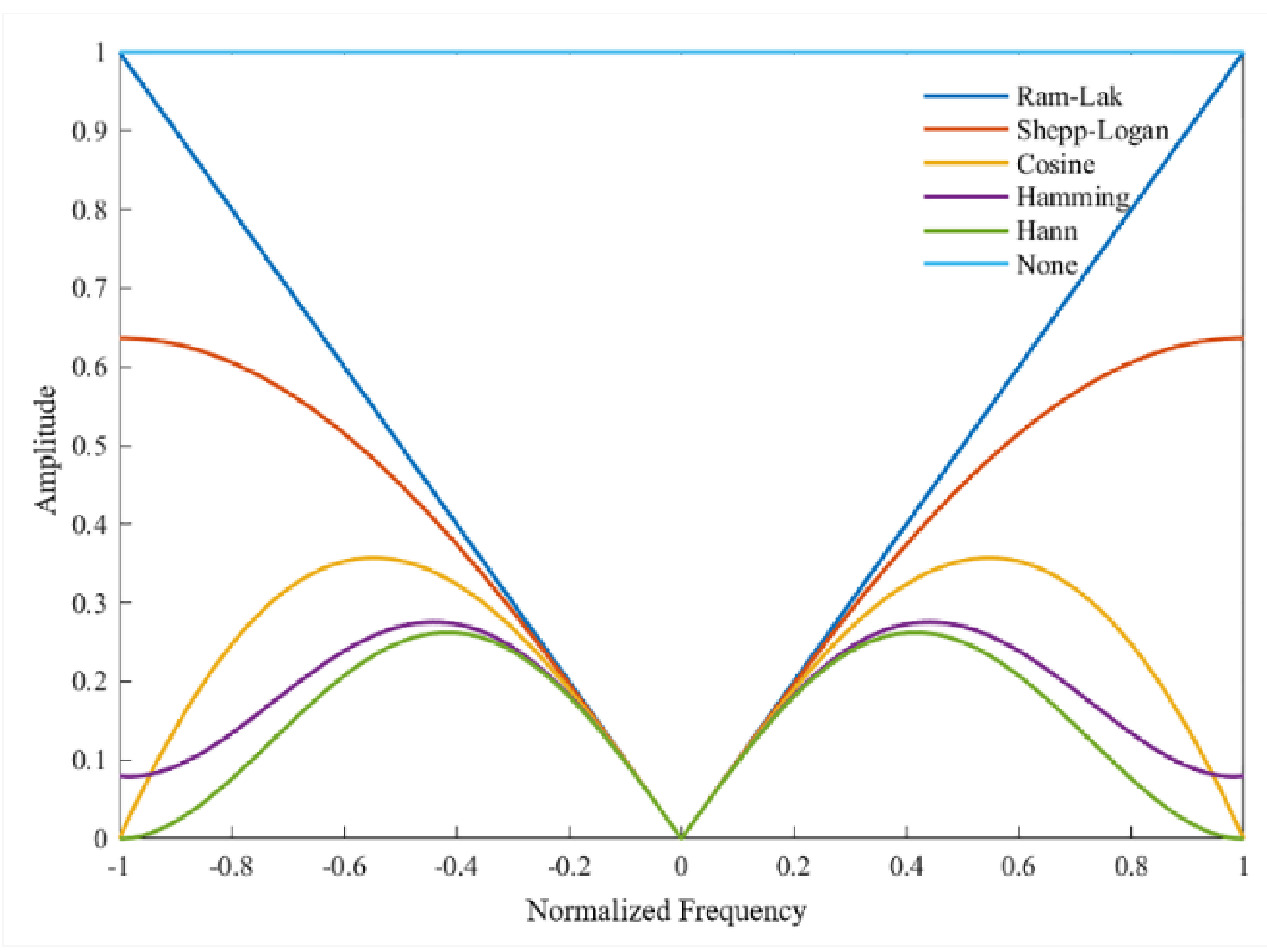

FIG. 1. Spectral characteristics of the filters critical to the unified frequency-domain model.

During the backprojection step, the filtered projection data $q(s_k,\theta)$ are defined at discrete detector positions $s_k$. To compute the backprojection value at image grid points $(x,y)$ ,it is necessary to interpolate the discrete $q(s_k,\theta)$ to continuous positions $s'$ based on the ray equation $s' = x\sin\theta + y\cos\theta$ [32]. The principles and characteristics of commonly used interpolation methods are described below.

Nearest interpolation is the simplest interpolation method. Let u be the normalized offset. The spatial kernel function of Nearest interpolation is[33]:

$$I_{\text{Nearest}}(u) = \begin{cases} 1, & \text{if} |u| \le 0.5 \\ 0, & \text{otherwise} \end{cases}, \tag{7}$$

Its frequency response (Fourier transform) is[34]:

$$\hat{I}_{\text{Nearest}}(\omega) = \text{sinc}(\omega) = \frac{sin(\pi\omega)}{\pi\omega}, \tag{8}$$

The spatial kernel function of Linear interpolation is[35]:

$$I_{\text{Linear}}(u) = \begin{cases} 1-|u|, & \text{if} |u| \le 1 \\ 0, & \text{otherwise} \end{cases}, \tag{9}$$

Its frequency response is[36]:

$$\hat{I}_{\text{Linear}}(\omega) = [\text{sinc}(\omega)]^2 = \left(\frac{sin(\pi\omega)}{\pi\omega}\right)^2, \tag{10}$$

The commonly used approximate spatial kernel function for cubic spline interpolation is[37]:

$$I_{\text{Spline}}(u) = \begin{cases} \frac{2}{3} - |u| + \frac{1}{2}|u|^3, & 0 \le |u| < 1 \\ \frac{1}{6}(2-|u|)^3, & 1 \le |u| < 2 \\ 0, & \text{otherwise} \end{cases}, \tag{11}$$

Its frequency response approximately satisfies [38]:

$$\hat{I}_{\text{Spline}}(\omega) \approx C[\text{sinc}(\omega)]^4 = C\left(\frac{sin(\pi\omega)}{\pi\omega}\right)^4 \tag{12}$$

Piecewise Cubic Hermite Interpolating Polynomial (Pchip) constructs a monotonicity-preserving cubic Hermite polynomial on each discrete interval $[s_k,$ $s_{k+1}]$ using the projection values $q(s_k,\theta)$,$q(s_{k+1},\theta)$ and estimated derivative values[39].

Therefore, the choice of interpolation method essentially represents a selection of frequency domain characteristics: whether to pursue a wide main lobe to preserve details (Spline), pursue low side lobes to suppress artifacts (Pchip), or pursue ultimate computational efficiency (Nearest).

Let $f(x,y)$ represent the reconstructed image, and $P(\omega,\theta)$ denote the Fourier transform [40] of the projection data at angle $\theta$. According to the projection slice theorem, the ideal reconstruction process in the continuous domain can be expressed as [41]:

$$f(x,y)=\frac{1}{2}\int_0^{\pi}\left[\int_{-\infty}^{\infty}|\omega|\,P(\omega,\theta)e^{i\omega(x\cos\theta+y\sin\theta)}d\omega\right]d\theta, \quad (13)$$

Based on the definition of the Fourier transform of the projection data, let $p(s,\theta)$ represent the original projection data, where $s$ is the perpendicular distance from the projection line to the coordinate origin. The Fourier transform of $P(\omega,\theta)$ is:

$$P(\omega,\theta)=\int_{-\infty}^{\infty}p\,(s,\theta)e^{-i2\pi\omega s}ds, \quad (14)$$

The filtered spectrum $Q(\omega,\theta)$is:

$$Q(\omega,\theta)=H(\omega)|\omega|P(\omega,\theta), \quad (15)$$

where $H(\omega)$ is the filter frequency response function (e.g., Ram-Lak, Hann, etc.), and the ramp factor $|\omega|$ compensates for the blurring effect introduced by the back-projection process.

Through the one-dimensional inverse Fourier transform, the filtered continuous projection signal$q(s,\theta)$ in the spatial domain is obtained:

$$q(s,\theta)=\int_{-\infty}^{\infty}Q\,(\omega,\theta)e^{i2\pi\omega s}d\omega, \quad (16)$$

In the back projection integral, it is necessary to calculate the value of $q(s',\theta)$ at position $s'$. Since $q(s,\theta)$is continuous in the spatial domain but only its sampled values $q(s_k,\theta)$ at discrete detector positions $s_k$ are available in practical operations, interpolation is required to obtain the value at $s'$, denoted as $\hat{q}(s',\theta)$ This can be expressed as:

$$\hat{q}(s',\theta) = \sum_k q\,(s_k,\theta) I\left(\frac{s'-s_k}{\Delta s}\right), \tag{17}$$

where $I$ is the normalized interpolation kernel function, and $\Delta s$ represents the detector element spacing.

Let $\theta_m$ denote the m-th projection angle and $M$ the total number of projection angles. The final reconstructed image $f(x,y)$ is obtained by performing back projection integration over all filtered and interpolated projection values $\hat{q}(s',\theta)$ across all projection angles $\theta$:

$$f(x,y) = \Delta\theta \sum_{m=0}^{M-1} \hat{q}\,(x\cos\theta_m + y\sin\theta_m\,, \theta_m), \tag{18}$$

where $\Delta\theta = \pi/M$ represents the angular sampling interval.

Integrating Equations (13) to (18), we obtain the reconstructed image expression accounting for the synergistic interaction between interpolation and filtering:

$$f(x,y) = \Delta\theta \sum_{m=0}^{M-1} \left[ \begin{array}{c} \sum_k \left( \int_{-\infty}^{+\infty} [H(\omega)|\omega| P(\omega,\theta_m)]\, e^{i2\pi\omega s_k} d\omega \right) \cdot \\ I\left(\frac{x\cos\theta_m + y\sin\theta_m - s_k}{\Delta s}\right) \end{array} \right] \tag{19}$$

In approximate analysis, the interpolation process of discrete samples $q(s_k,\theta)$ yielding $\hat{q}(s',\theta)$ can be regarded as a discrete convolution, whose frequency response approximately equals the product of the discrete sequence spectrum $Q(\omega,\theta)$ and the interpolation kernel spectrum $\hat{I}(\omega)$ Combined with the filtered spectrum $Q(\omega,\theta) = H(\omega)|\omega|P(\omega,\theta)$ and considering discrete sampling effects[42], Eq.(18) can be approximately expressed as:

$$f(x,y) \approx \Delta\theta \sum_{m=0}^{M-1} F^{-1} \left\{ \int_{-\infty}^{+\infty} [H(\omega)|\omega| P(\omega,\theta_m)]\, \hat{I}(\omega) e^{i2\pi\omega s} d\omega \right\} \Big|_{s = x\cos\theta_m + y\sin\theta_m} \tag{20}$$

where $F^{-1}$ denotes the inverse Fourier transform. This expression holds approximately under ideal conditions where the detector sampling is sufficiently dense and the discrete convolution effects of the interpolation kernel are negligible. $\hat{I}(\omega)$ is the Fourier transform of the interpolation kernel function, and different interpolation methods correspond to different morphological characteristics: if $\hat{I}(\omega)$ has a narrow main lobe, it preserves more high-frequency components; if $\hat{I}(\omega)$ has low sidelobes, it reduces aliasing artifacts and improves smoothness. For example, Eq. (8) shows that Nearest interpolation exhibits severe high-frequency oscillations, amplifying noise and introducing block artifacts; Eq. (10) indicates that Linear interpolation causes mid-frequency attenuation and partial detail loss; Eq. (12) demonstrates that Spline interpolation features a wide main lobe and low sidelobes, preserving effective bandwidth while suppressing artifacts.

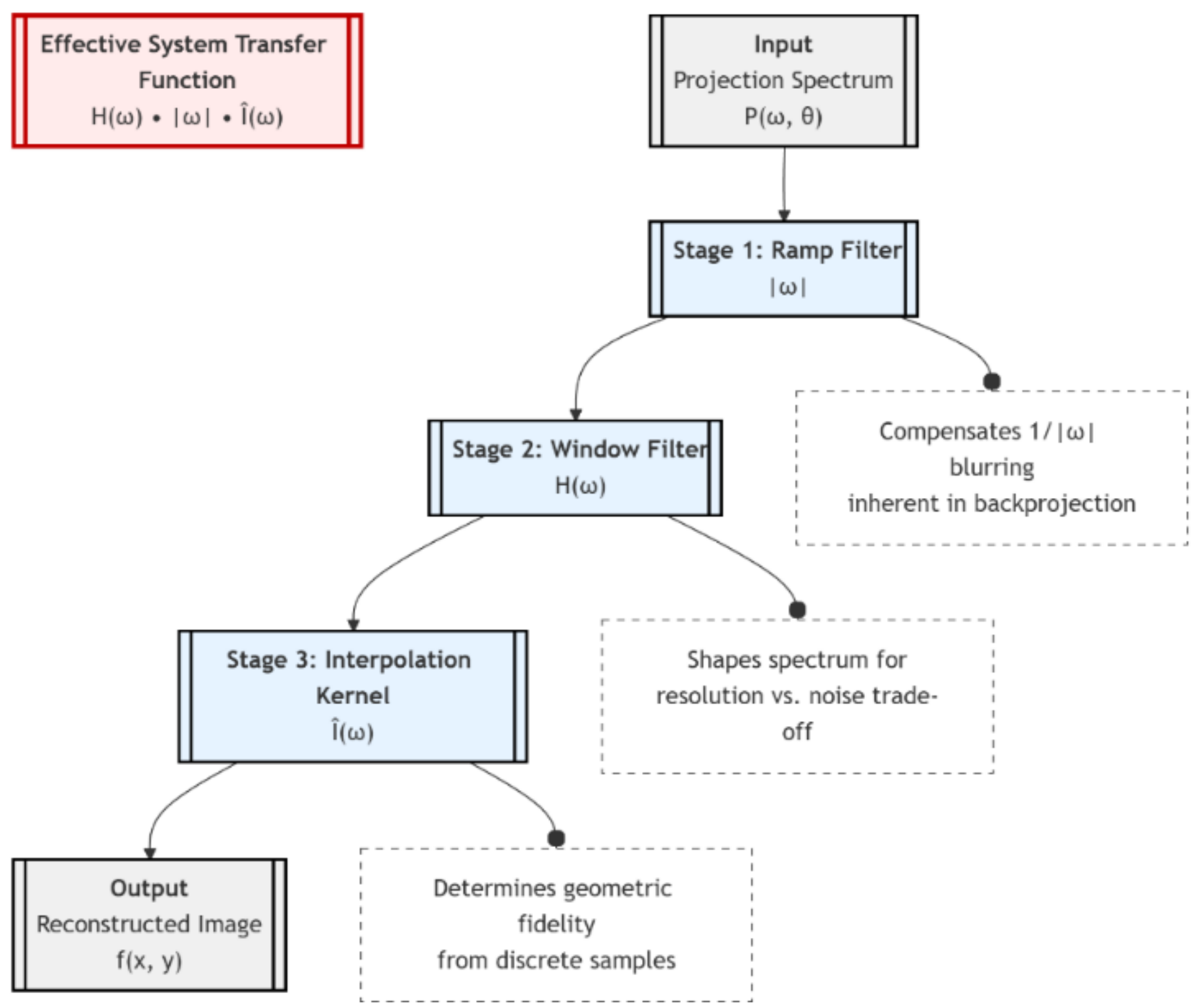

FIG. 2. Block diagram of the unified cascaded modulation model for filtered backprojection reconstruction.

Through the approximate model of Eq. (20), two key functional components can be clearly identified:

Filtering component $H(\omega)|\omega|P(\omega,\theta_m)$: Shapes the frequency domain of the original projection spectrum $P(\omega,\theta_m)$. The ramp factor $|\omega|$ compensates for back projection blurring, while the window function $H(\omega)$ suppresses noise (e.g., Hann filter) or preserves details (e.g., Ram-Lak filter) according to specific requirements.

Interpolation component $\hat{I}(\omega)$: Operates on the filtered spectrum, aiming to approximate the spectral characteristics of the continuous signal $q(s,\theta)$ by shaping the frequency domain through $\hat{I}(\omega)$ after discrete sampling $q(s_k,\theta)$, thereby correcting spectral aliasing distortions introduced by discrete sampling.

Eq. (20) constitutes our proposed unified frequency-domain model for FBP reconstruction. The derived model reveals that the FBP reconstruction process is effectively a cascaded linear system (Fig. 2). The first stage, defined by $H(\omega)|\omega|$, performs a mandatory deblurring and application-specific spectral weighting. The second stage, governed by$\hat{I}(\omega)$, compensates for the aliasing and smoothing introduced by discrete sampling. The overall quality is thus determined by the product of these two frequency responses, $\boldsymbol{H}(\omega)|\omega|\boldsymbol{P}(\omega,\theta_m)$ , which acts as the effective transfer function of the entire FBP pipeline.

We adopt the structural similarity index (SSIM), peak signal-to-noise ratio (PSNR), and reconstruction time as quantitative metrics for comparing the interpolation-filter combinations. The formulas for SSIM and PSNR are shown in Eqs. (21) - (22), respectively:

$$M_{\mathrm{SSI}} = \frac{(2\mu_x\mu_y+c_1)(2\sigma_{xy}+c_2)}{(\mu_x^2+\mu_y^2+c_1)(\sigma_x^2+\sigma_y^2+c_2)}, \tag{21}$$

$$R_{\mathrm{PSN}} = 10\,log_{10}\left[\frac{max^2(f)}{\frac{1}{N}\sum_{i=1}^{N}(x_i-y_i)^2}\right] \tag{22}$$

where $x$ and $y$ represent the two images, $\mu_x$ and $\mu_y$ are the mean of the images and respectively, $\sigma_x^2$ and $\sigma_y^2$ are the variance of the images, $\sigma_{xy}$ their covariance; $f$ denotes the reference (original) image, $max(f)$ its maximum possible pixel value, $x_i$ and $y_i$ the corresponding pixel values in the reference and reconstructed images, respectively, $N$ the total number of pixels in the image. The reconstruction time is measured as the total computational time consumed from the start of the reconstruction algorithm to the complete output of the reconstructed image.

To quantify the performance differences among the parameter combinations, we employed a standard Shepp-Logan phantom. Projection data comprising 180 views from 0° to 179° were reconstructed using all 20 combinations of four interpolation methods (Nearest, Linear, Spline, Pchip) and five filters (Ram-Lak, Shepp-Logan, Cosine, Hamming, Hann). Each combination underwent five independent trials, and the results were averaged. Computational efficiency was assessed using a high-resolution 2048×2048-pixel phantom, with the mean reconstruction time for each combination recorded over ten runs. Three combinations were applied to real cable CT data from the Shanghai Synchrotron Radiation Facility (SSRF). A sample was scanned at 30 keV, acquiring 540 projections over 180° with 200 ms exposure. From the 600×2048-pixel projections, 600 cross-sectional slices were reconstructed at 1448×1448 pixels, and the average per-slice reconstruction time was recorded for each combination.

As quantified in Table 1, the PSNR values are predominantly determined by the interpolation method, with higher-order methods (Spline, Pchip) consistently outperforming lower-order ones (Linear, Nearest) for any given filter. This aligns with the model's prediction that the interpolation kernel's frequency response $\hat{I}(\omega)$ acts as a passband filter on the filtered spectrum. The supremacy of the Spline + Ram-Lak combination (highest PSNR) is a direct validation of the model: the Ram-Lak filter's uncompromising high-frequency emphasis ($H(\omega)$) is faithfully preserved by the Spline kernel's wide main lobe ($\hat{I}(\omega)$), resulting in a system response $H(\omega) \cdot \hat{I}(\omega)$ that effectively recovers fine details. The SSIM metrics further validate the model's predictive power for perceptual image quality. The optimal performance of Spline + Hann arises from a different type of synergy: the Hann filter's effective suppression of high-frequency noise and Gibbs ringing, combined with the Spline kernel's ability to minimize interpolation-induced artifacts, produces a spectrum conducive to structural integrity. While our model focuses on the frequency-domain determinants of image quality, computational cost is a critical practical consideration. The reconstruction time analysis completes the holistic evaluation by quantifying the computational cost of each combination. The data reveals a clear trade-off: combinations employing high-order interpolation (e.g., Spline) deliver superior image fidelity but at a significantly higher computational cost, while simpler methods (e.g., Nearest) offer supreme efficiency for real-time processing.

TABLE I. Rrediction validation (PSNR, SSIM, and reconstruction time) results under noise-free conditions.

| Method | Metric | Nearest | Linear | Spline | Pchip |
|---|---|---|---|---|---|
| Ram-Lak | PSNR | 26.87 | 27.38 | 28.34 | 27.98 |
| | SSIM | 0.641 | 0.866 | 0.840 | 0.838 |
| | Time(s) | 0.082 | 0.173 | 1.156 | 1.171 |
| Shepp-Logan | PSNR | 26.58 | 26.78 | 27.67 | 27.36 |
| | SSIM | 0.703 | 0.902 | 0.888 | 0.885 |
| | Time(s) | 0.080 | 0.170 | 1.154 | 1.161 |
| Cosine | PSNR | 25.55 | 25.58 | 26.19 | 26.02 |
| | SSIM | 0.796 | 0.943 | 0.945 | 0.941 |
| | Time(s) | 0.078 | 0.160 | 1.140 | 1.158 |
| Hamming | PSNR | 24.88 | 24.85 | 25.35 | 25.23 |
| | SSIM | 0.826 | 0.946 | 0.951 | 0.948 |
| | Time(s) | 0.081 | 0.163 | 1.141 | 1.158 |
| Hann | PSNR | 24.65 | 26.64 | 25.08 | 24.98 |
| | SSIM | 0.833 | 0.947 | 0.953 | 0.950 |
| | Time(s) | 0.083 | 0.168 | 1.164 | 1.171 |

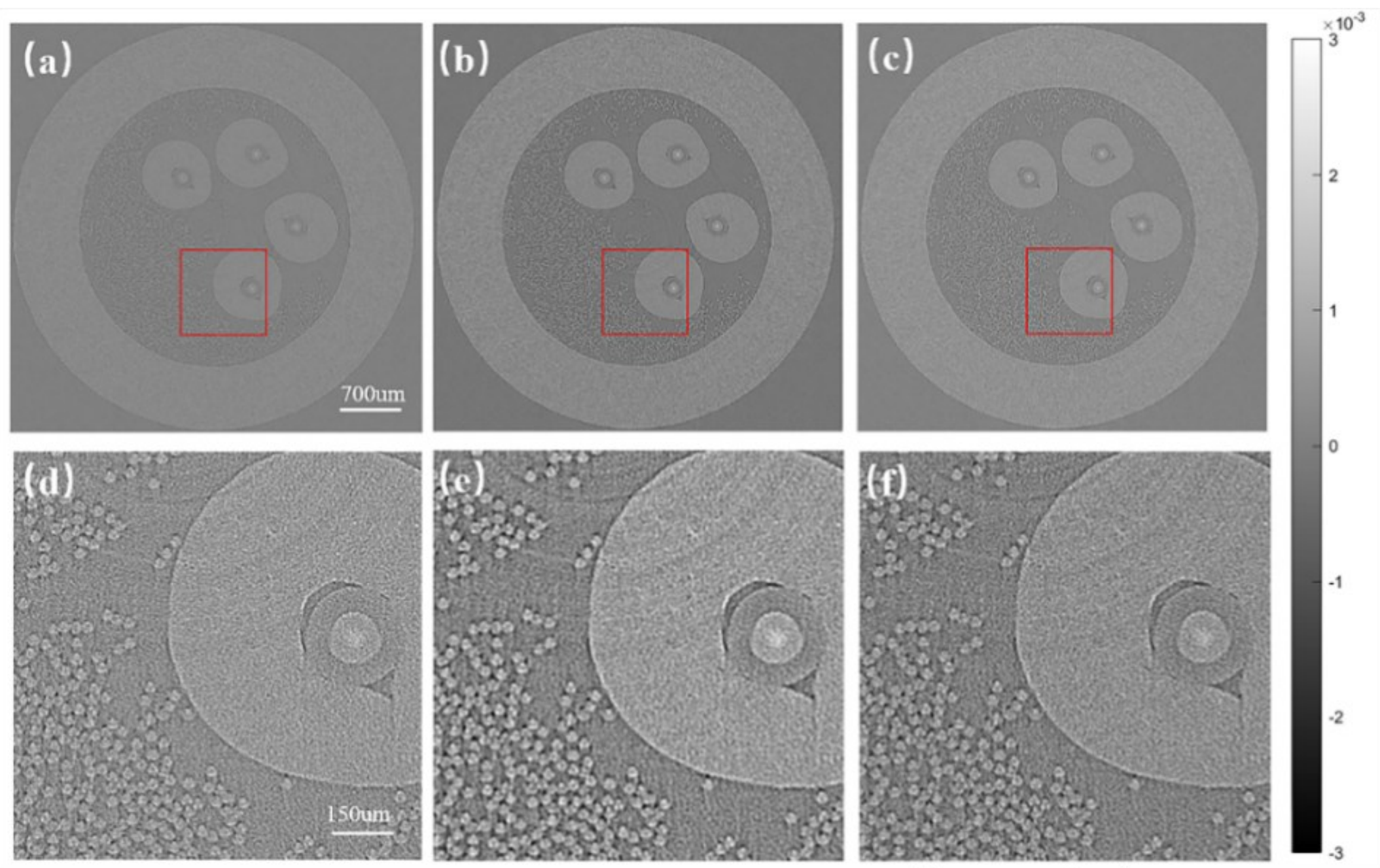

FIG. 3. Experimental validation on real cable CT data using different FBP combinations. (a, d) Spline + Ram-Lak. (b, e) Spline + Hann. (c, f) Nearest + Cosine. Panels (d)-(f) show magnified views of the corresponding red-box regions in (a)-(c), respectively.

To assess the generalizability of our model from numerical phantoms to practical scenarios, we applied the three combinations to real synchrotron cable CT data (Fig. 3). The results confirm the predictive power of our model under realistic imaging conditions. For the real cable data (Fig. 3a, d), Spline + Ram-Lak resolves fine details and sharp interfaces, such as the conductor-insulator boundary. This result confirms that the high-frequency emphasis of the Ram-Lak filter, when paired with Spline interpolation, is effective for reconstructing authentic material boundaries and achieving high quantitative accuracy. The Spline + Hann result (Fig. 3b, e) shows effective noise suppression and homogeneous regions, yielding a coherent representation of the overall structure. This demonstrates the utility of this combination for maintaining structural integrity in the presence of real-world noise and imperfections, consistent with the model's emphasis on the filter's role in noise control. The reconstruction using Nearest+ Cosine (Fig. 3c, f) shows pronounced stairstep artifacts along the cable edges and internal interfaces. This loss of geometric fidelity is a characteristic outcome of the nearest-neighbor interpolation method, providing a clear example of how a physically antagonistic pair can degrade image quality in practical applications, regardless of the object being imaged.

The experimental results across numerical phantoms and real-world data consistently validate the predictions of the unified frequency-domain model. The quantitative metrics (PSNR, SSIM, time) and qualitative visual outcomes all confirm that the synergistic or antagonistic interaction between $\boldsymbol{H}(\omega)$ and $\hat{\boldsymbol{I}}(\omega)$ is the principal determinant of reconstruction performance. The striking consistency between the numerical phantom and the real-world synchrotron data demonstrates the generalizability of the model beyond simulated conditions, confirming its value for real-world tomographic imaging applications.

The experimental results consistently confirm that the performance of FBP reconstruction is not merely a function of independently optimized components but is fundamentally governed by the synergistic interaction between the frequency responses of the filter and the interpolation kernel. This work establishes that this interaction can be formally modeled as a cascaded modulation process, providing a unified physical framework that transcends the traditional paradigm of isolated parameter tuning. Our cascaded model of $H(\omega) \cdot \hat{I}(\omega)$ provides the missing theoretical link that explains why certain combinations excel in specific tasks.

Although validated in parallel-beam CT, our cascaded modulation model generalizes to broader linear inverse problems. Future work will extend it to complex geometries like cone-beam CT, evaluate its robustness to strong nonlinearities such as metal artifacts, and examine how GPU acceleration reshapes the observed CPU-based performance hierarchy.

In conclusion, this work establishes a unified frequency-domain model that moves beyond traditional FBP parameter selection by conceptualizing filter-interpolation interaction as a cascaded modulation process. This provides not only a predictive framework for tomographic reconstruction but also a principled approach to linear inverse problems, demonstrating how algorithmic synergy grounded in physical principles offers a powerful pathway for advancing computational imaging systems. The model thus serves as a compelling case study in the principled physical engineering of linear inverse problems, providing a concluding perspective from tomography to computational physics.

## ACKNOWLEDGMENTS

We thank the staff members of X-ray Imaging and Biomedical Applications Beamline (https://cstr.cn/31124.02.SSRF.BL13HB) at Shanghai Synchrotron Radiation Facility (SSRF) for providing technical support and assistance in data collection and analysis.

## AUTHOR DECLARATIONS

The authors have no conflicts to disclose.

## DATA AVAILABILITY

The data that support the findings of the study are available on request from the corresponding author.